\documentclass[%
 reprint,
showpacs,
 amsmath,amssymb,
 aps, 
prstab,
]{revtex4-1}

\usepackage{graphicx}
\usepackage{subfigure}
\usepackage{dcolumn}
\usepackage{multirow}
\usepackage{bm}
\usepackage{overpic}
\usepackage{mathrsfs}
\usepackage{xcolor}
\usepackage{siunitx}

\begin{document}


\title{Plasma Lenses for Relativistic Laser Beams in Laser Wakefield Accelerators}

\author{Ming Zeng}
\email{ming.zeng@desy.de}
\affiliation{
 Deutsches Elektronen-Synchrotron DESY, 22607 Hamburg, Germany
}
\author{Alberto Martinez de la Ossa}
\affiliation{
 Deutsches Elektronen-Synchrotron DESY, 22607 Hamburg, Germany
}
\author{Kristjan Poder}
\affiliation{
 Deutsches Elektronen-Synchrotron DESY, 22607 Hamburg, Germany
}
\author{Jens Osterhoff}
\affiliation{
 Deutsches Elektronen-Synchrotron DESY, 22607 Hamburg, Germany
}%

\date{\today}

\begin{abstract}
Focusing petawatt-level laser beams to a variety of spot sizes for different applications is expensive in cost, labor and space. In this paper, we propose a plasma lens to flexibly resize the laser beam by utilizing the laser self-focusing effect. Using a fixed conventional focusing system to focus the laser a short distance in front of the plasma, we can adjust the effective laser beam waist within a certain range, as if a variety of focusing systems were used with the plasma lens acting as an adjustable eyepiece in a telescope. Such a setup is a powerful tool for laser wakefield accelerator experiments in state-of-art petawatt laser projects and allows for scanning focal spot parameters.
\end{abstract}

\maketitle


\section{\label{sec:intro}Introduction}

Nowadays petawatt lasers have become high-priority tools for studying the intrinsic properties of the microscopic physical world~\cite{SEL}. Manipulation of such powerful lasers is a big challenge due to the lack of high damage-threshold optical materials.  The current solution is to use large beam apertures so that the laser power is spread across a large area of the optical element to prevent damage. For example, $\rm SiO_2$ has a damage threshold of the order of \SI{1}{J/cm^2} if it is irradiated by a femtosecond laser~\cite{LGallaisJAP2015}. If we assume the laser pulse duration is \SI{50}{fs}, the wavelength is \SI{800}{nm} and the laser beam is perfectly Gaussian, to focus a \SI{1}{PW} laser beam a mirror with a \SI{0.34}{m} diameter is required to prevent mirror damage, resulting in high costs for high-quality focusing optics. The cost issue is compounded by the fact that for each laser system multiple focusing systems are required for different applications: short focal length optics for laser-solid interactions and long focal length ones for laser-gas interactions. A particularly demanding application is the laser wakefield accelerator (LWFA), where the laser beam should be focused to a matched spot size that stabilizes laser propagation in the plasma, with the size of this matched spot changing with laser power, the plasma density, and the plasma channel depth in the external guiding case~\cite{EEsareyRMP2009, CBenedettiPOP2012, WLuPRSTAB2007, IKostyukovPOP2004, KPoderPPCF2018}. For photon-nuclear interaction applications a changeable laser spot size is also advantageous for maximizing the electron flux~\cite{MZengPST2017}. 

In order to avoid damaging laser optics, the beam near field diameter $D$ must scale as $D \propto \sqrt{P}$ with laser power $P$. The f-number $N \equiv f/D \propto k w_0$, where $f$ is the focal length, $w_0$ is the laser beam waist at focus and $k = 2\pi / \lambda$ is the wavenumber for laser with wavelength $\lambda$. Thus $f \propto w_0 \sqrt{P} \propto P/a_0$, where $a_0 \approx 8.5\times 10^{-10} \lambda\left[{\rm \mu m}\right]\sqrt{I_0\left[{\rm W/cm^2}\right]} \propto \sqrt{P}/w_0$ is the normalized laser amplitude and $I_0$ is the peak intensity. For LWFA studies an optimum $a_0$ is required even with increasing $P$~\cite{WLuPRSTAB2007}, resulting in the focal length $f \propto P$ being extremely long. For example, focal lengths on the order of 10 meters are required for \SI{1}{PW} lasers~\cite{WPLeemansPRL2014}, while for \SI{100}{PW} lasers the focal lengths would be on the order of 1000 meters. With such scales presenting obvious difficulties, a focusing system with a variable focal spot size and small footprint is urgently required.

\begin{figure}
	\subfigure{\includegraphics[width=0.47\textwidth]{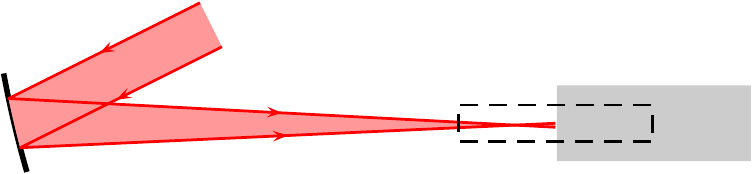}}\\
	\vspace{0.005\textheight}
	\subfigure{
    	\begin{overpic}[width=0.47\textwidth]{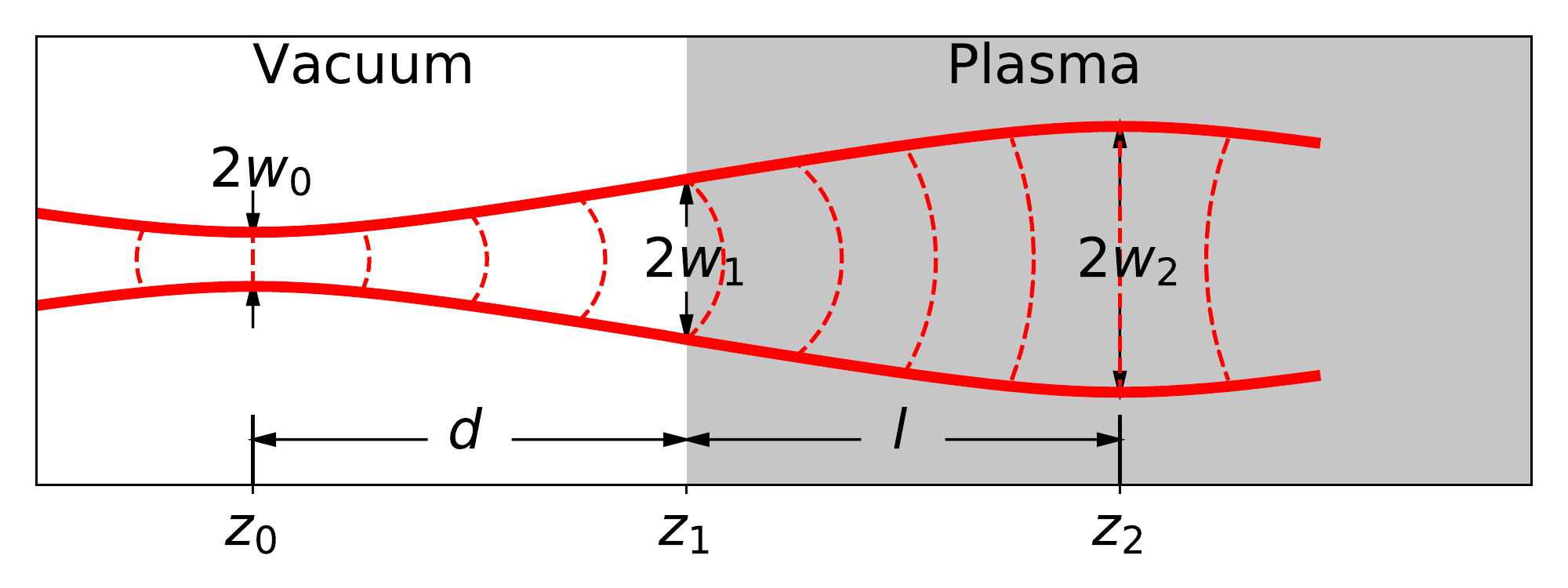}
        	\put(2,35){\line(5,1){58}}
        	\put(98,35){\line(-1,1){12}}
		\end{overpic}}
	\caption{Illustration of the plasma lens. The laser is focused in vacuum to $z=z_0$ with waist of $w_0$ by a conventional optical system and then enters a plasma at $z=z_1$. The plasma can thereafter reshape the wavefront so that the wavefront becomes flat again at $z=z_2$ with the laser size $w=w_2$. In the bottom subplot, the solid red curves illustrate the transverse envelope, and the dashed red curves illustrate the wavefront of the laser beam.}
   \label{fig:ref_illu}
\end{figure}

In this paper, we introduce such a focusing system employing the laser self-focusing effect in plasmas~\cite{CEMaxPRL1974}. The scheme is illustrated in Fig.~\ref{fig:ref_illu}. A high power laser beam is pre-focused to $z_0$ by a conventional focusing system with a focal length of $f_0$ to a spot size of $w_0$. After propagating a distance $d$, the laser enters the plasma region starting at $z_1$. After another distance $l$, the laser beam is refocused to a spot size $w_2$ at $z_2$ because of the self-focusing effects~\cite{CEMaxPRL1974}. Consequently the total function of this system is to focus the laser to a spot size of $w_2$ within a distance of $L = f_0 + d +l$. Due to the strong plasma response, $L$ can be much shorter than a conventional focusing system resulting in the same spot size. In addition, $w_2$ is adjustable by changing $d$, $l$ and the plasma density $n_p$.

This paper is organized as follows. In Sec.~\ref{sec:analytic} an analytical model describing the plasma lens is developed. In Sec.~\ref{sec:empirical} an empirical model is found based on particle-in-cell simulations. The empirical and analytical model are compared in Sec.~\ref{sec:compare} and a full scale LWFA simulation employing the proposed plasma lens is presented in Sec.~\ref{sec:full_scale}.

\section{\label{sec:analytic}Weakly Relativistic Self-refocusing model}

The evolution of the laser pulse in an underdense plasma in the long pulse and slow profile variation assumptions is given by~\cite{PMoraPOP1997}
\begin{eqnarray}\label{eq:tilde_a}
  \left(\nabla_{\perp}^2-i2k\partial_z\right)\tilde{a}=n\left(\frac{1}{\gamma}-1\right)\tilde{a},
\end{eqnarray}
where $z$ is the laser propagation distance, $n$ is the local plasma electron density and $\gamma$ is the plasma electron Lorentz factor. Normalized units are adopted with densities being normalized to the background plasma density $n_p$, wavenumbers to the plasma wavenumber $k_p = \sqrt{4\pi r_e n_p}$ where $r_e \approx 2.82 \times 10^{-15}\ \rm m$ is the classical electron radius, and lengths normalized to the plasma skin depth $k_p^{-1}$. The cylindrically symmetrical normalized laser vector potential $\tilde{a}$, including the transverse phase modulation but excluding the laser quiver factor $\exp\left(-ikz+i\omega t\right)$, is given by
\begin{eqnarray}
  \tilde{a}=a\exp{\left(i u r^2\right)} \exp{\left(-\frac{r^2}{w^2}\right)}, \label{eq:tilde_a_form}
\end{eqnarray}
where $r$ is radius, $a=a\left(z\right)$ is the axial normalized laser vector potential amplitude, $u=u\left(z\right)$ is the spatial phase modulation factor (effectively, the radius of the wavefront curvature is $k/2u$), and $w=w\left(z\right)$ is the laser spot size.

In general, solving Eq.~(\ref{eq:tilde_a}) analytically is difficult. However, it has been found that under weakly relativistic assumptions where the plasma density is unperturbed $n = 1$, the approximate solution of Eq.~(\ref{eq:tilde_a}) for a linearly polarized laser beam can be obtained by calculus of variations~\cite{DAndersonPOF1979, ZMShengJOSAB1996, PSprangleIEEE1987} as the following (derivation is given in App.~\ref{sec:eq_spot_size}):
\begin{eqnarray}
  \left|a\right|^2 w^2&=&\left|a_0\right|^2 w_0^2 \label{eq:energy_conserve}\\
  \dfrac{d^2 w}{dz^2} &=& \dfrac{4}{k^2 w^3} \left( 1-\dfrac{\left|a_0\right|^2w_0^2}{32} \right), \label{eq:w}
\end{eqnarray}
where Eq.~(\ref{eq:energy_conserve}) represents the energy conservation law and Eq.~(\ref{eq:w}) describes the evolution of the laser spot size in the weakly relativistic regime $a \lesssim 1$~\cite{KCTzengPRL1998}.

We apply Eq.~(\ref{eq:w}) to our case with the initial conditions at the vacuum-plasma interface ($z=z_1$) reading
\begin{eqnarray} 
  w_1 \equiv \left. w\right|_{z_1} &=& w_0 \sqrt{1+\dfrac{d^2}{z_R^2}}\\
  \left. \dfrac{d w}{dz} \right|_{z_1} &=& \dfrac{w_0^2 d}{z_R^2 w_1}, \label{eq:w_initial}
\end{eqnarray}
where $z_R=k w_0^2 / 2$ is the Rayleigh length. After integrating Eq.~(\ref{eq:w}) the refocused spot size can be found by requiring $\mathrm{d}w/\mathrm{d}z=0$ and is given by
\begin{eqnarray}
  w_2 \equiv \left. w\right|_{z_2} &=& w_0 \sqrt{1+\dfrac{d^2}{z_R^2} \cdot \dfrac{1}{1 -\left(1 + \dfrac{d^2}{z_R^2}\right)\dfrac{32}{a_0^2 w_0^2}}} \label{eq:w_2}.
\end{eqnarray}
The length of the plasma lens $l$ is given by
\begin{equation}
    l \equiv z_2-z_1 =\dfrac{d}{\dfrac{a_0^2 w_0^2}{32}\left(1 + \dfrac{d^2}{z_R^2}\right)^{-1}-1} \label{eq:l}
\end{equation}
with the limit
\begin{eqnarray}
  d < z_R \sqrt{\dfrac{a_0^2 w_0^2}{32}-1} \equiv d_{\rm M}. \label{eq:d_threshould}
\end{eqnarray}
Thus $d_{\rm M}$ is the upper limit of the pre-focusing distance $d$ for self-refocusing to occur.

Eqs.~(\ref{eq:w_2}), (\ref{eq:l}) and (\ref{eq:d_threshould}) show that in the analytical model the effective laser spot size $w_2$ and the plasma lens thickness $l$ are the functions of the normalized laser vector potential amplitude at vacuum focus $a_0$, the vacuum focal size $w_0$, the vacuum Rayleigh length $z_R$ and the pre-focusing distance $d$ (all the lengths are normalized to $k_p^{-1}$). Effectively, $w_2$ and $l$ are functions of $k$ (normalized to $k_p$), $w_0$, $a_0$ and $d$, indicating the parameter space to be scanned in the simulation studies in Sec.~\ref{sec:empirical}.

\section{\label{sec:empirical}Simulations and Empirical Formulas}

The analytic model presented in Sec.~\ref{sec:analytic} is valid for $a_0 \lesssim 1$, while for realistic LWFA applications a relativistic laser intensity is required. To examine the behaviour of the plasma lens in the highly relativistic regime,  three-dimensional (3D) particle-in-cell (PIC) simulations using the code OSIRIS~\cite{OSIRIS} were performed. To characterize the plasma lens, the refocused spot size $w_2$ and plasma lens length $l$ were found by scanning a four-dimensional parameter space of ($k$, $w_0$, $a_{0\max}$, $d$), where $a_{0\max}$ is the peak normalized laser amplitude along the laser co-moving coordinate $\xi=z-ct$ (i.~e.\ at the vacuum focus, $a_0$ is a function of $\xi$, varies between 0 and $a_{0\max}$). The longitudinal profile of the laser pulse is a bell shape (both the rise and fall envelopes of the laser take the form $10 X^3 - 15 X^4 + 6 X^5$ in the range $0\leq X \leq 1$ where $X=\left|\xi - \xi_0\right| / \tau$ and $\xi_0$ is the pulse center), thus the averaged $a_0$ is approximately $a_{0\max}/2$. The initial full-width-at-half-maximum pulse duration was $\tau_\mathrm{FWHM}=4$ (time is normalized to $\omega_p^{-1} = k_p^{-1} c^{-1}$). The simulations were performed using a moving window of length 10, co-moving with the laser pulse towards the positive $z$ direction with the speed of light. The transverse extent of the simulation box was adjusted to be between $10w_2$ and $12w_2$. In order to ensure convergence, two resolutions for each of the simulations were used as shown in Tab.~\ref{tab:resolutions}. Simulation time steps were set close to the Courant condition to prevent numerical dispersion. The number of macro-particles per cell is 8 and they are initiated with a thermal momentum of $p_{\rm t}/m_e c=0.01$.

\begin{table}
\begin{center}
  \begin{tabular}{ c  c  c }
    \hline
    $k$ & Low resolution & High resolution \\
    \hline
    10 & $256\times 256^2$ & $512\times 512^2$ \\
    20 & $512\times 256^2$ & $1024\times 512^2$ \\
    30 & $512\times 256^2$ & $1024\times 512^2$ \\
    40 & $1024\times 256^2$ & $2048\times 512^2$ \\
    \hline
  \end{tabular}
  \caption{Longitudinal and transverse numbers of cells used for different values of $k$ in the simulations.}
  \label{tab:resolutions}
\end{center}
\end{table}

\begin{figure}
   \includegraphics[width=0.48\textwidth]{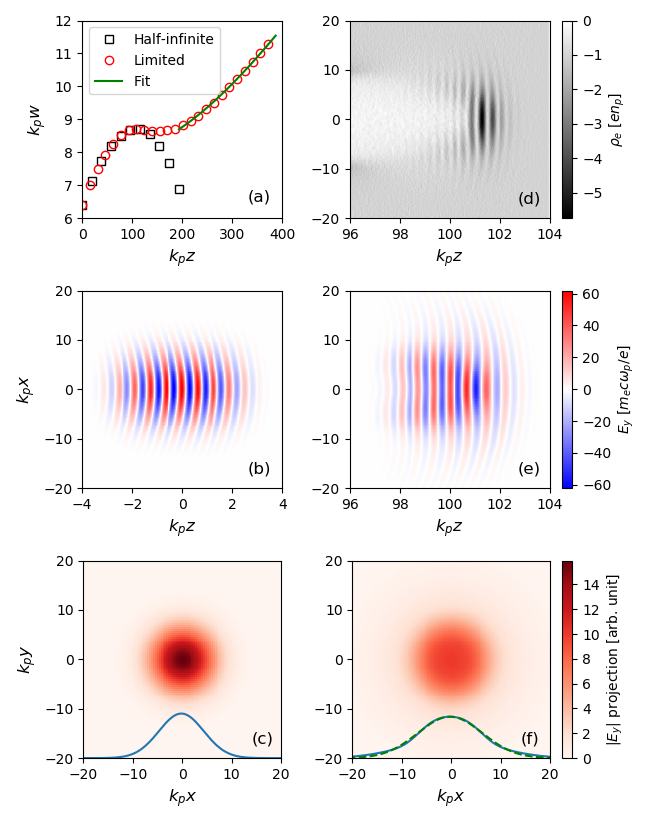}
   \caption{Example PIC simulations with $k=10$, $w_0=4$, $a_{0\max}=10$, $d=100$ and plasma region at $z>0$ (thus the laser is pre-focused at $z=-100$). (a) Spot size $w$ as a function of propagation distance $z$, for a half infinite plasma starting from $z=0$ (black squares) and a limited plasma region from $z=0$ to $z=z_2$ (red circles), with $z_2=104\pm 4$ in this case. The green curve is a Gaussian beam propagation fit for the limited plasma for $z>200$. (b) Side view (slice at $y=0$) of $E_y$ at the beginning of the simulation. (c) Front view (projection to the $x$-$y$ plane) of $|E_y|$ at the beginning of the simulation. (d) Side view of the plasma electron charge density at $z\approx z_2$. (e) Side view of $E_y$ at $z\approx z_2$. (f) Front view of $|E_y|$ at $z\approx z_2$. The solid blue lines in (c) and (f) are the line projections of the front views, and the dashed green line in (f) is the Gaussian fit of the blue solid line, showing that the blue solid line slightly deviates from a Gaussian profile.}
   \label{fig:ref_example}
\end{figure}

For each simulation, snapshots of the absolute value of the laser electric field are projected to the $x$-$y$ plane (front view). Two-dimensional Gaussian fits are then performed to obtain the spot size $w$ for each snapshot. Example simulation results are shown in Fig.~\ref{fig:ref_example}. A laser with $k=10$ and $a_{0\max}=10$ is focused to $w_0=4$ at $z=-100$. It then enters the plasma region starting at $z=0$ and self-refocusing occurs. As seen in Fig.~\ref{fig:ref_example} (a), $w$ reaches a local maximum $w_2=8.7$ at $z=z_2=104\pm 4$ for the half-infinite plasma case, plotted as black squares. In the case of a limited plasma region ending at $z=z_2$ which is plotted as red circles in Fig.~\ref{fig:ref_example} (a), after the plasma ends the laser spot size approximately evolves as
\begin{eqnarray}
  w=w_{0e}\sqrt{1+\dfrac{\left(z-z_{0e}\right)^2}{z_{Re}^2}}, \label{eq:z_w_vacuum}
\end{eqnarray}
where $w_{0e}$ is the effective focal size, $z_{0e}$ is the effective focal position and $z_{Re}=k w_{0e}^2/2$ is the effective Rayleigh length with $k=10$ in this case. The evolution of the spot size in the vacuum of the rear side can be fitted by Eq.~(\ref{eq:z_w_vacuum}), resulting in $z_{0e} = 53.8$ and $w_{0e}=8.0$. The data in the region $z<200$ are ignored in the fit, because the far-distance measurement can better represent the beam size due to the non-perfectly Gaussian profile of the laser beam. The mismatch $z_{0e}\neq z_2$ and $w_{0e}\neq w_2$ is due to the laser transverse profile not being perfectly Gaussian at $z=z_2$, as highlighted in Fig.~\ref{fig:ref_example} (e) and (f). However, with only a $10\%$ difference between the spot sizes, $w_2$ is used as the effective focal size of our plasma lens in the following.

\begin{figure}
   \includegraphics[width=0.48\textwidth]{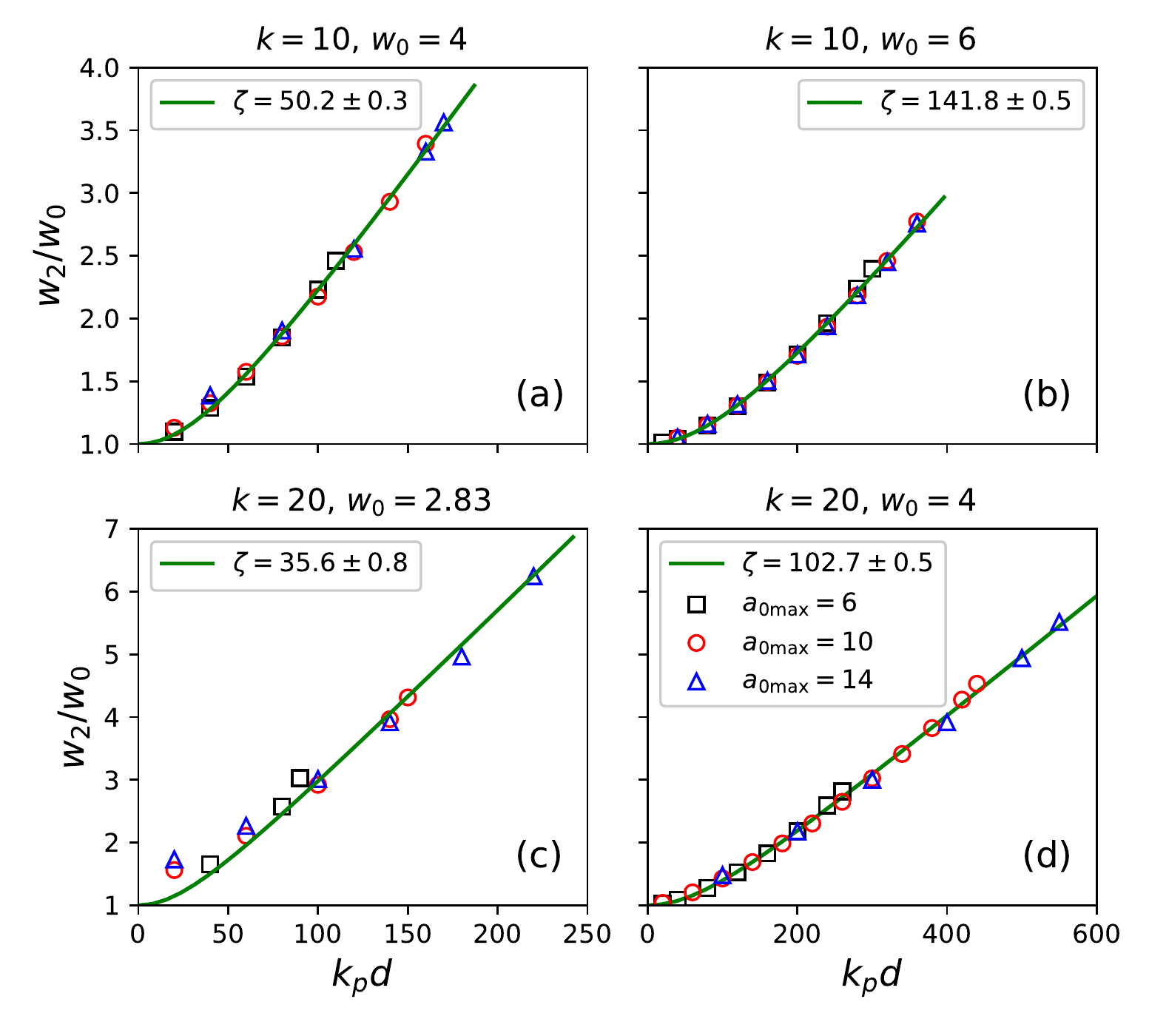}
   \caption{The effective focal size - vacuum focal size ratio $w_2/w_0$ vs.\ the distance from the vacuum focus to the plasma $d$ for $k=10$ (top) and $k=20$ (bottom). In each of the subplots, $k$ and $w_0$ are fixed, while $a_0$ varies from $6$ (black squares), $10$ (red circles) to $14$ (blue triangles). The green lines are the fits by Eq.~(\ref{eq:d_w2}), with the parameter $\zeta$ shown in the legends.}
   \label{fig:d_w2}
\end{figure}

\subsection{\label{ssec:d_w2}Effective focal size vs.\ pre-focusing distance}

The dependence of the ratio $w_2/w_0$ on the pre-focusing distance $d$ for varying laser strengths $a_{0\max}$, wavenumbers $k$ and initial spot sizes $w_0$ is shown in Fig.~\ref{fig:d_w2}. It can be seen that $w_2/w_0$ and $d$ approximately obey the relation
\begin{eqnarray}
  \dfrac{w_2}{w_0}=\sqrt{1+\dfrac{d^2}{\zeta^2}}, \label{eq:d_w2}
\end{eqnarray}
where $\zeta$ is a parameter which depends on $k$ and $w_0$, but only weakly on $a_{0\max}$. $\zeta$ can be regarded as a modified Rayleigh length to be discussed in Sec.~\ref{sec:compare}. As is evident from Fig~\ref{fig:d_w2} (c), when $d$ is very small the behaviour of $w_2/w_0$ deviates from the general trend. This is caused by large cavitation, occurring when $a>w^2$ as described in Appx.~\ref{sec:cavitation}.

\begin{figure}
   \includegraphics[width=0.48\textwidth]{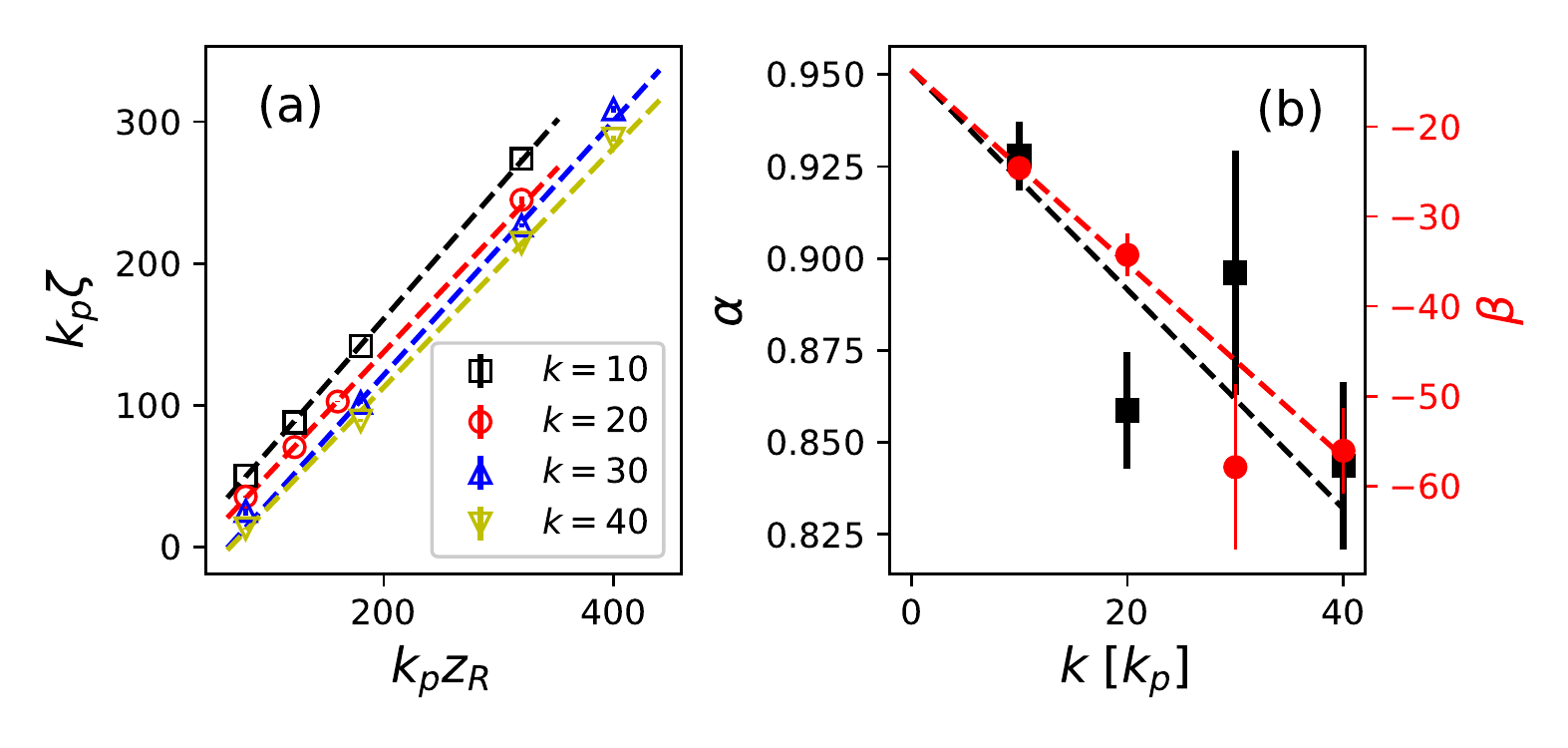}
   \caption{(a) The modified Rayleigh length $\zeta$ as a function of vacuum Rayleigh length $z_R$ with $k=10$ (black squares), $20$ (red circles), $30$ (blue triangles) and $40$ (yellow inverted triangles). (b) The linear fit parameters $\alpha$ (black squares) and $\beta$ (red dots) vs.\ $k$, where the fit function is $\zeta=\alpha z_R + \beta$. The dashed lines are the linear fits, and the vertical lines are the error bars.}
   \label{fig:zR_vs_zeta}
\end{figure}
A weak dependence of $\zeta$ on $a_{0\max}$ appears over a large parameter range. Although it may be not intuitively understandable, such a weak dependence is a natural requirement for self-refocusing to occur. Due to the laser pulse having a limited duration, $a_0$ changes along its temporal profile with the center having a maximum value of $a_{0\max}$, whereas $a_0$ is smaller at the front and rear of the pulse. Were $\zeta$ to have a strong dependence on $a_0$, self-refocusing of different temporal slices would be different, rendering self-refocusing of the whole pulse impossible. In other words, only in the regime where such weak dependence is satisfied, can self-refocusing be observed.

The variation of $\zeta$ with $z_R$ and $k$ is plotted in Fig.~\ref{fig:zR_vs_zeta} (a). For a fixed value of $k$ a linear dependence of $\zeta$ on $z_R$ is observed. Thus the data in Fig.~\ref{fig:zR_vs_zeta} (a) can be fitted with $\zeta=\alpha z_R + \beta$, with the resulting fit parameters $\alpha$ and $\beta$ shown in Fig.~\ref{fig:zR_vs_zeta} (b)~\footnote{In this paper, all the errors from the fits are the standard deviations if not explicitly specified}. The dependence of $\alpha$ and $\beta$ on $k$ is again found from a linear fit as
\begin{eqnarray}
  \alpha &=& (-0.0027 \pm 0.0017)k + (0.950\pm 0.028), \label{eq:alpha} \\
  \beta &=& (-1.17 \pm 0.20)k + (-12.6 \pm 2.7)
\end{eqnarray}
The expressions for $\alpha$ and $\beta$ then allow for an empirical expression for $\zeta$ to be obtained:
\begin{eqnarray}
  \zeta \approx 0.95 z_R - 1.2k - 13, \label{eq:zeta_empirical}
\end{eqnarray}
where $k$-dependence of $\alpha$ is neglected, because for $k\leq 40$ the first term on the RHS of Eq.~(\ref{eq:alpha}) is less than 10\% of the second term.
 
\subsection{\label{ssec:d_w0_l}Plasma lens thickness vs.\ pre-focusing distance}

\begin{figure}
   \includegraphics[width=0.48\textwidth]{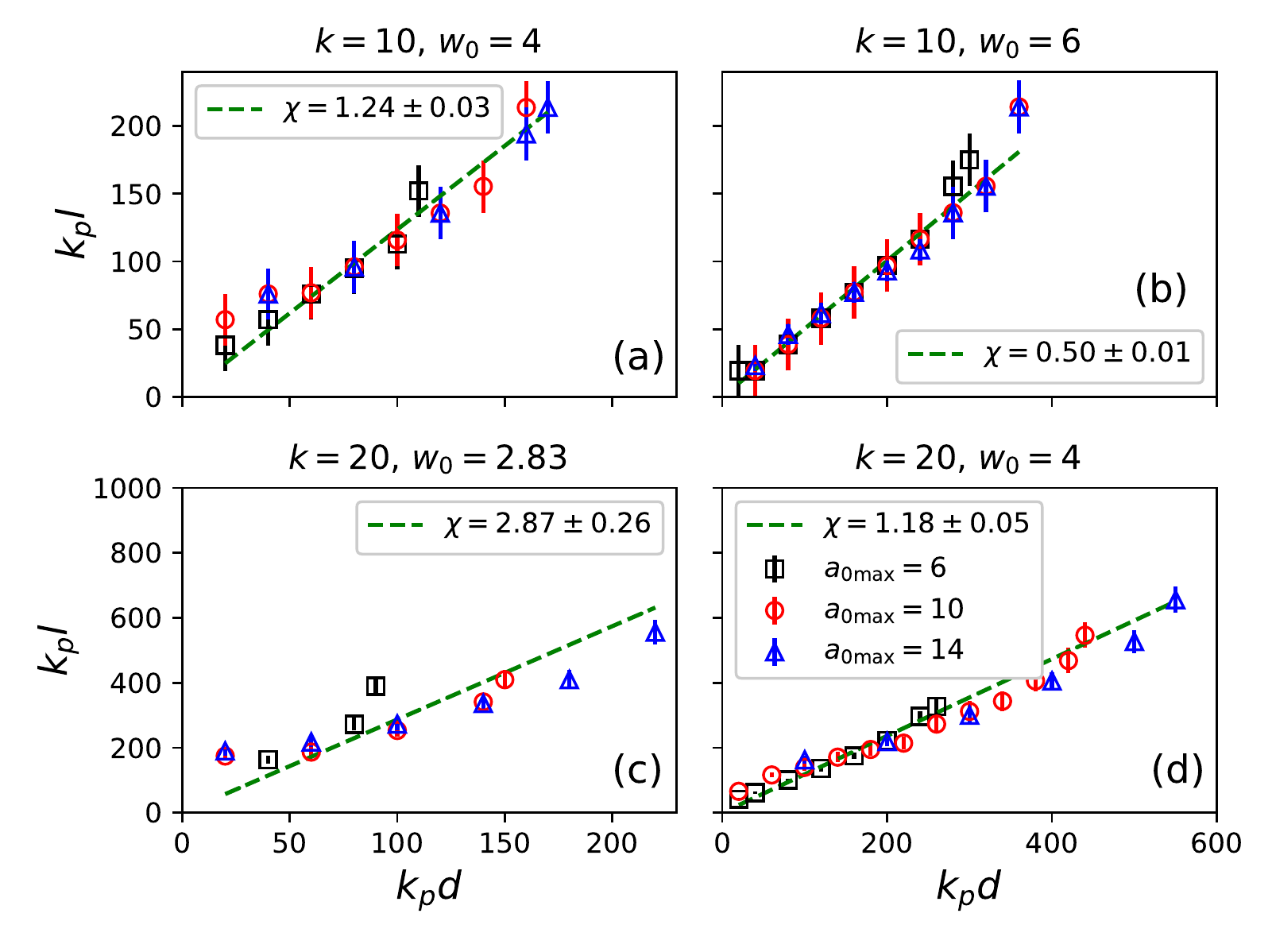}
   \caption{Plasma lens thickness $l$ as a function of the pre-focusing distance $d$ for different values of $k$, $w_0$ and $a_{0\max}$, and $a_{0\max}=6$ (black squares), 10 (red circles) and 14 (blue triangles). The vertical lines are error bars due to the simulation dumping intervals. The green dashed lines are fits by Eq.~(\ref{eq:l_chi_d}), with the slope $\chi$ shown in the legends.}
   \label{fig:d_vs_l}
\end{figure}

Some examples of plasma lens thickness $l$ as a function of the pre-focusing distance $d$ are plotted in Fig.~\ref{fig:d_vs_l}. It is evident that $l$ is almost proportional to $d$ for a fixed $k$ and $w_0$. Some exceptions are seen in Fig.~\ref{fig:d_vs_l} (c) for small values of $d$, again due to large cavitation occurring when $a>w^2$ as described in Appx.~\ref{sec:cavitation}. The variation of plasma lens thickness $l$ with $d$ is modelled with the relation
\begin{eqnarray}
  l=\chi d, \label{eq:l_chi_d}
\end{eqnarray}
where $\chi$ is a fit parameter depending on both $k$ and $w_0$.

\begin{figure}
   \includegraphics[width=0.32\textwidth]{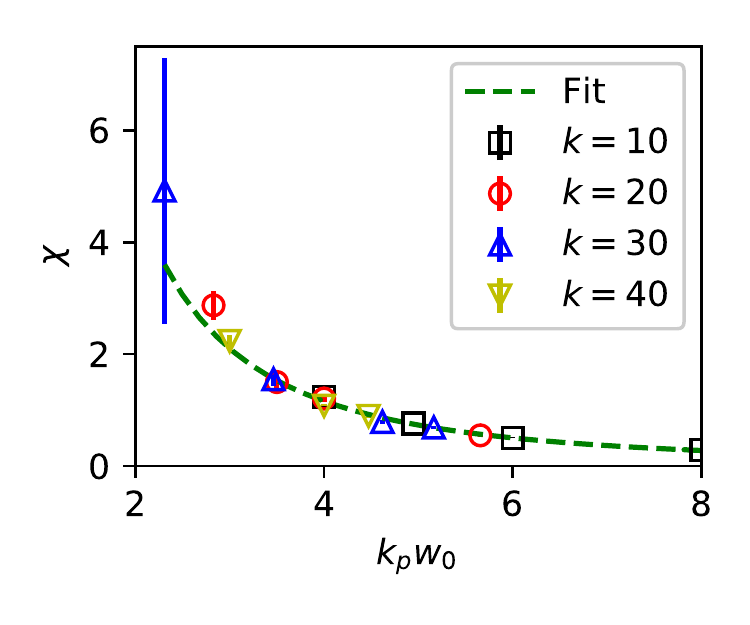}
   \caption{The ratio $\chi = l/d$ as a function of the vacuum focal size $w_0$ for $k=10$ (black squares), 20 (red circles), 30 (blue triangles) and 40 (yellow inverted triangles). The green dashed line is the fit of all the data by Eq.~(\ref{eq:chi_w0}).}
   \label{fig:w0_vs_chi}
\end{figure}

The variation of $\chi$ with the vacuum spot size $w_0$ is then plotted in Fig.~\ref{fig:w0_vs_chi}. As can be seen,  all data points lie approximately on the same curve regardless of the value of $k$. An expression for $\chi$ can now be written as
\begin{eqnarray}
  \chi = \mu w_0^{\nu} \label{eq:chi_w0}
\end{eqnarray}
and performing a fit to the data presented in Fig.~\ref{fig:w0_vs_chi} results in $\mu = 21.0 \pm 1.8$ and $\nu = -2.08 \pm 0.05$. Thus
\begin{eqnarray}
  \chi \approx 21.0 w_0^{-2.08} \label{eq:chi_empirical}
\end{eqnarray}
and the final empirical formula for the plasma lens thickness $l$ is obtained as
\begin{eqnarray}
  l \approx 21.0\dfrac{d}{w_0^{2.08}}. \label{eq:l_d_empirical}
\end{eqnarray}

\subsection{\label{ssec:dM}Limitation of the pre-focusing distance}

\begin{figure}
   \includegraphics[width=0.34\textwidth]{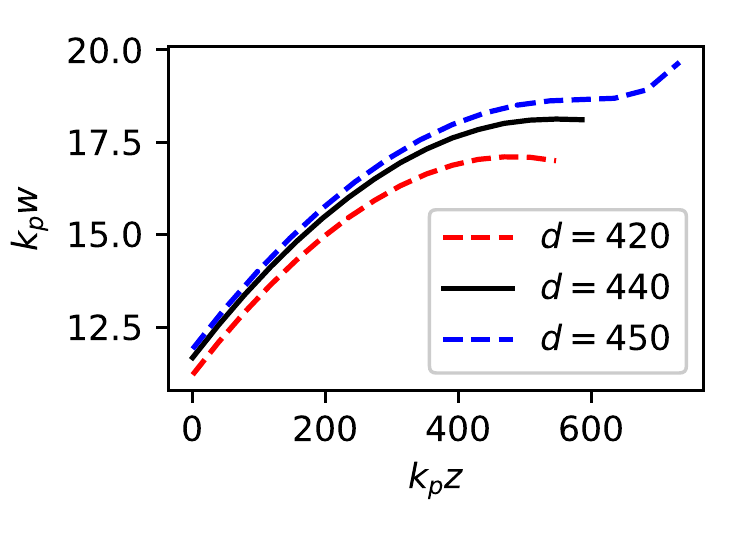}
   \caption{Examples of laser spot size evolution, showing the threshold of $d$ for self-refocusing to occur. The parameters are $k=20$, $w_0=4$ and $a_{0\max}=10$. The three curves are for $d=420$ (red dashed line), 440 (black solid line) and 450 (blue dashed line).}
   \label{fig:dM}
\end{figure}

Although in all of our simulations the laser powers were chosen to be higher than the critical powers for self-focusing ($P/P_c = a_0^2 w_0^2/32 >1$), self-refocusing does not always occur because the initial conditions in our case $\left. dw/dz \right|_{z_1}>0$ may lead to a continuous growth of $w$. An example of self-refocusing not occurring is shown in Fig.~\ref{fig:dM}, where $k$, $w_0$ and $a_{0\max}$ are fixed while $d$ changes from 420 to 450. It can be seen that for $d=420$ and 440 the local maxima of $w$ are present while for $d=450$ the curve does not have a local maximum. Thus the threshold value of $d$ for self-refocusing to occur is approximately 445, or $d_{\lim} = 445 \pm 5$ in this case.

\begin{table}
\begin{center}
  \begin{tabular}{ c | c | c | c | c || c | c | c | c | c }
    \hline
    $k$ & $w_0$ & $a_{0\max}$ & $d_{\lim}$ ($\pm 5$) & $d_{\rm M}$ & $k$ & $w_0$ & $a_{0\max}$ & $d_{\lim}$ ($\pm 5$) & $d_{\rm M}$ \\
    \hline
    \multirow{6}{*}{10} & \multirow{3}{*}{4} & 6 & 115 & 150 & \multirow{6}{*}{20} & \multirow{3}{*}{2.83} & 6 & 105 & 89 \\
    & & 10 & 165 & 271 & & & 10 & 155 & 183 \\
    & & 14 & 175 & 388 & & & 14 & 225 & 268 \\
    \cline{2-5}\cline{7-10}
    & \multirow{3}{*}{6} & 6 & 305 & 544 & & \multirow{3}{*}{4} & 6 & 265 & 299 \\
    & & 10 & 365 & 937 & & & 10 & 445 & 542 \\
    & & 14 & 365 & 1324 & & & 14 & 595 & 776 \\
    \hline
  \end{tabular}
  \caption{Comparisons of the threshold for refocusing obtained from simulations $d_{\lim}$ and from the analytical model $d_{\rm M}$.}
  \label{tab:dM}
\end{center}
\end{table}

In Tab.~\ref{tab:dM}, some values of the threshold for refocusing obtained from simulation results $d_{\lim}$ are compared with the analytical results $d_{\rm M}$ from Eq.~(\ref{eq:d_threshould}). One can see that in some of the cases, especially for $a_{0\max}=6$, $d_{\lim}$ agrees with $d_{\rm M}$ reasonably well, while in the other cases $d_{\rm M} > d_{\lim}$. Finding the general expression for $d_{\lim}$ is computationally expensive, thus we take $d_{\rm M}$ to represent $d_{\lim}$.

\subsection{\label{ssec:tau}Influence of pulse duration}

In the previous discussion we took the averaged information of the laser temporal slices. The slice differences can be small if the laser beam has a long pulse duration. However, in short pulse cases the slice differences may be important. Thus in the following we discuss pulse duration effects.

\begin{figure}
   \includegraphics[width=0.48\textwidth]{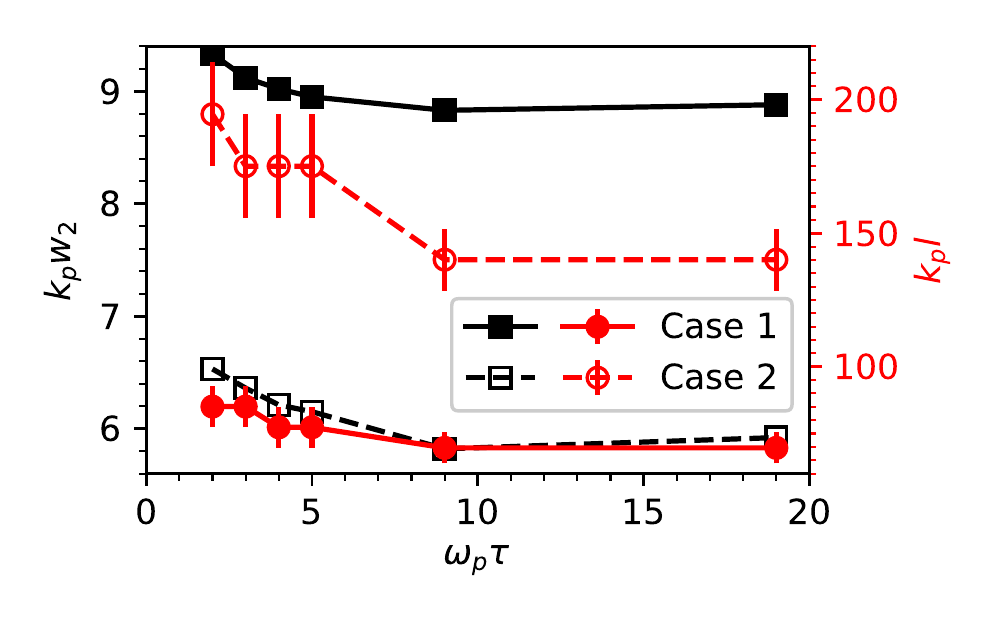}
   \caption{The effective focal size $w_2$ and the plasma lens thickness $l$ vs.\ the laser pulse duration $\tau$ for two cases: $k=10$, $w_0=6$, $a_{0\max}=14$, $d=160$ (solid lines and solid markers) and $k=20$, $w_0=3.5$, $a_{0\max}=10$, $d=100$ (dashed lines and hollow markers). The black color is for $w_2$, the red color is for $l$, and the vertical lines on the red markers are the error bars due to the simulation dumping intervals.}
   \label{fig:w2_l_vs_tau}
\end{figure}

Figure~\ref{fig:w2_l_vs_tau} shows the influence of changing the initial pulse duration $\tau$ for two example cases. It is evident that with $\tau$ increasing from 2 to 19, both $w_2$ and $l$ decrease. The influence of $\tau$ is stronger at smaller values of $\tau$. This is due to a shorter laser pulse creating a higher gradient in the plasma electron density, which then has a higher impact on the laser evolution. In a longer laser pulse case, only the front of the laser experiences a high electron density gradient, while the majority of the laser pulse is in a relatively longitudinally uniform density region. This is the physical reason behind the effect of changes of $\tau$ becoming less severe for longer values of the pulse length.

Although the refocusing effect changes with pulse duration $\tau$, the resulting variation of $w_2$ is less than \SI{10}{\percent} for all cases. The variation of $l$ is large, but less influential as explained in the following. Because $dw/dz$ approaches $0$ while $w$ approaches $w_2$, the change of $w$ near the maximum is too small to ensure an exact measurement of $l$. Thus measuring of $l$ itself has a large systematic error. Meanwhile, because $l$ is usually much smaller than the dephasing length as shown in Sec.~\ref{sec:full_scale}, knowing the exact value of the plasma lens thickness $l$ is less important.

In all the simulations of the previous subsections we have chosen $\tau=4$, where $w_2$ and $l$ approximately take the median values. Thus we consider our former conclusions reliable for a wide range of $\tau$, i.~e.\ $\tau \geq 2$.

\section{\label{sec:compare}Comparison of the Simulation Results, Analytical model and the Empirical Formulas}
By comparing Eqs.~(\ref{eq:w_2}) and (\ref{eq:d_w2}), an analytical correspondence of $\zeta$ can be written as
\begin{eqnarray}
  \zeta_{a} = z_R \sqrt{1 -\left(1 + \dfrac{d^2}{z_R^2}\right)\dfrac{32}{a_0^2 w_0^2}}. \label{eq:zeta_a}
\end{eqnarray}
This suggests $\zeta_a$ is mainly dependent on, but smaller than the Rayleigh length $z_R$, which is similar to the empirical expression for $\zeta$ in Eq.~(\ref{eq:zeta_empirical}). Also by comparing Eqs.~(\ref{eq:l}) and (\ref{eq:l_chi_d}) an analytical correspondence of $\chi$ can be found as
\begin{eqnarray}
  \chi_{a} = \dfrac{1}{\dfrac{a_0^2 w_0^2}{32}\left(1 + \dfrac{d^2}{z_R^2}\right)^{-1}-1}. \label{eq:chi_a}
\end{eqnarray}
In the limit of the laser power being much higher than the critical power for self-focusing $a_0^2 w_0^2/32 \gg 1$~\cite{PSprangleIEEE1987}, applying the weakly relativistic approximation $a_0 \lesssim 1$ (thus $w_0 \gg 1$) and with the assumption of short distance from vacuum focus to plasma $d \ll z_R$, Eq.~(\ref{eq:chi_a}) reduces to $\chi_a \propto w_0^{-2}$, which is again similar to the empirical expression for $\chi$ in Eq.~(\ref{eq:chi_empirical}).

\begin{figure}
   \includegraphics[width=0.48\textwidth]{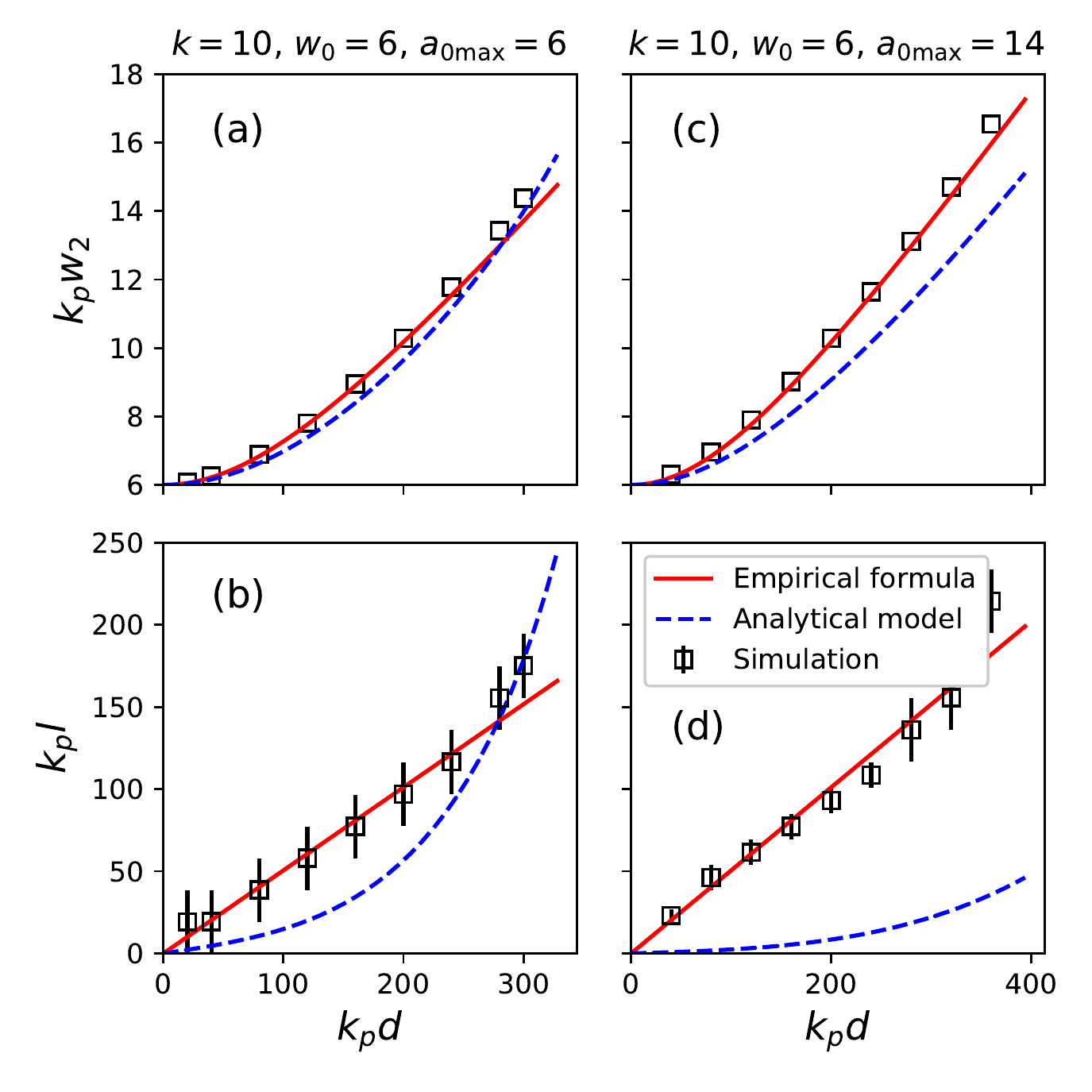}
   \caption{Comparison of empirical formulas Eqs.~(\ref{eq:d_w2}), (\ref{eq:zeta_empirical}) and (\ref{eq:l_d_empirical}) (red solid lines), the analytical model Eqs.~(\ref{eq:w_2}) and (\ref{eq:l}) (blue dashed lines), with the example simulation results (black squares). The parameters are $k=10$, $w_0=6$ and $a_{0\max}=6$ (left) or 14 (right).}
   \label{fig:compare}
\end{figure}

The variation of the refocused spot size $w_2$ and the plasma lens thickness $l$ with pre-focusing distance $d$ in both the empirical formulas and the analytical model (by setting $a_0=a_{0\max}/2$) is compared to some simulations in Fig.~\ref{fig:compare}. It is evident that in most cases the empirical formulas fit the simulations well, but the analytical model only partially agrees with the simulations for a smaller $a_{0\max}$. The disagreement arises from the fact that for $a_{0\max} \gg 1$, the plasma cannot be regarded as unperturbed and thus the weekly relativistic assumption used in the analytical model is no longer applicable.

\section{\label{sec:full_scale}Full Scale LWFA Simulations}
In order to show the physical scales of the plasma lens explicitly, unnormalized units will be used in this section. Using Eqs.~(\ref{eq:d_w2}), (\ref{eq:zeta_empirical}), (\ref{eq:l_d_empirical}), and applying the matching condition~\cite{WLuPRSTAB2007} at $z_2$ (i.~e.\ $k_p w_2 = 2 \sqrt{a_2}$), A plasma lens is designed for a \SI{1}{PW}, \SI{800}{nm} laser with \SI{108}{fs} pulse duration. 
A linear density ramp from $z=-20 k_p^{-1}= \SI{-216}{\micro\metre}$ to $z=0$ is followed by a density plateau with density $n_p = \SI{2.43e17}{\per\centi\metre\cubed}$ (thus $k_p^{-1} = \SI{10.8}{\micro\metre}$ and $\omega_p^{-1} = \SI{35.9}{fs}$). The laser frequency is thus $\omega/\omega_p = k/k_p = 84.7$, the FWHM pulse duration is $\tau = 3\omega_p^{-1}$ and the laser vacuum focus is set to $z=-80 k_p^{-1} = \SI{-864}{\micro\metre}$ with $w_0 = 2 k_p^{-1} = \SI{21.6}{\micro\metre}$, thus $a_0 = 8$. The influence of the density transition region is neglected, thus the pre-focusing distance is $k_p d = 80$. According to Eqs.~(\ref{eq:d_w2}), (\ref{eq:zeta_empirical}) and (\ref{eq:l_d_empirical}), this setup results in $w_2 = 4 k_p^{-1} = \SI{43.2}{\micro\metre}$ and $a_2=4$, with plasma lens thickness $l = 397 k_p^{-1} = \SI{4.29}{\milli\metre}$.

\begin{figure*}
   \includegraphics[width=2.0\columnwidth]{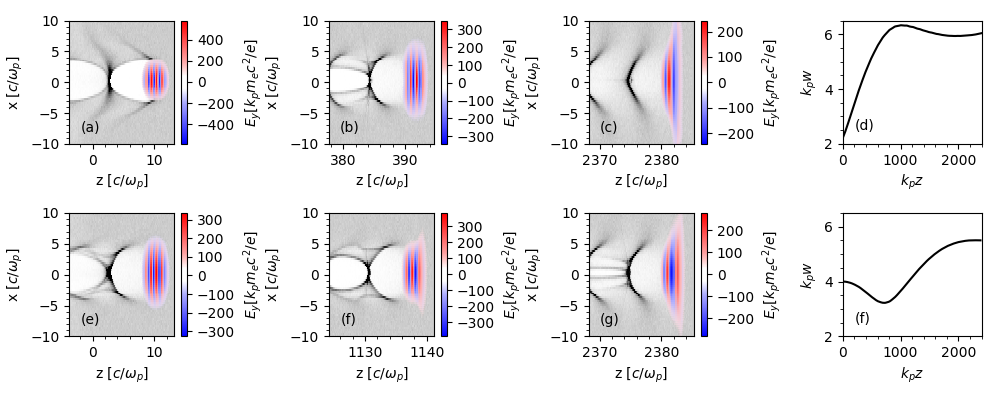}
   \caption{Two 2-cm-long 3D PIC simulations showing the LWFAs with or without a plasma lens. Top: with plasma lens. The laser beam has $k_p w_0 = 2$ and focused at $z=-80 k_p^{-1}$. Lu's matching condition is satisfied at $z=397 k_p^{-1}$. Bottom: without plasma lens. The laser beam has $k_p w_0 = 4$ and focused at $z=0$. Lu's matching condition is satisfied at $z=0$. The pseudo-color plots show the side views of the laser beams and the plasma densities, and the right-most plots show the laser beam front-view size evolution.}
   \label{fig:os_PT3D}
\end{figure*}

A confirmation simulation is run with the parameters just discussed, setting the simulation box size to $17 k_p^{-1} \times 40 k_p^{-1} \times 40 k_p^{-1}$ and number of cells to $4096 \times 256 \times 256$. The time step is $\Delta t = 4.147\times 10^{-3} \omega_p^{-1}$. Snapshots of a transverse slice through the laser pulse and laser spot size evolution are shown in the top panels of Fig.~\ref{fig:os_PT3D}. It is evident that at $z = l = 397 k_p^{-1}$, the laser size $w$ is increased to $4 k_p^{-1}$ and the bubble radius is also approximately $4 k_p^{-1}$. At  later times the laser front becomes not well guided and at around $z=2000 k_p^{-1}$ the spot size $w$ stabilizes at about $5.9 k_p^{-1}$. The source of the mismatch between the local maximum of the spot size $w$ and the prediction from the empirical formulas can be the change of pulse duration; $\omega_p\tau=3$ is used here while the value  $\omega_p\tau=4$ was used for obtaining the empirical formulas. Another source of error can be the extrapolation of the results for $10 \leq k/k_p \leq 40$ to $k/k_p=84.7$. Nevertheless, the bubble radius remains at about $4 k_p^{-1}$ for later times. Although the plasma lens parameters are chosen to satisfy Eq.~(\ref{eq:d_sf}) to avoid self-injection, a small bunch of electrons with a charge of \SI{16}{pC} is self-injected due to bubble evolution. However, the charge of this bunch is negligible compared to the loading capacity of this bubble which is a few nanocoulombs according to Eq.~(10) of Ref.~\cite{WLuPRSTAB2007} and Eq.~(10) of Ref.~\cite{MTzoufrasPRL2008}. Moreover, According to Eqs.~(3) and (4) of Ref.~\cite{WLuPRSTAB2007}, both the pump depletion length $l_{\rm dep} \approx \left( k/k_p \right)^2 \omega_p \tau k_p^{-1} = 2.15 \times 10^4 k_p^{-1} = \SI{232}{mm}$ and the dephasing length $l_d \approx \frac{2}{3}\left( k/k_p \right)^2 w_2 = 1.91 \times 10^4 k_p^{-1} = \SI{206}{mm}$ are much larger than $l$, thus the plasma lens does not significantly shorten the acceleration length.

To highlight the performance of the plasma lens, a simulation without the plasma lens was performed with $a_0=4$, $w_0 = 4k_p^{-1}=\SI{43.2}{\micro\metre}$ and $k_p d =0$. The results of this simulation are shown in the bottom panels in Fig.~\ref{fig:os_PT3D}. The spot size evolution, shown in Fig.~\ref{fig:os_PT3D}~(d) and Fig.~\ref{fig:os_PT3D}~(f) for the simulation with and without the plasma lens, respectively, is evidently not the same for the two cases. Nevertheless, the laser spot size approaches $k_pw=5.5$ in the case without the plasma lens, which is similar to the value of $k_pw=5.9$ for the simulation with the plasma lens. The spot size does not become stabilized at the matched value 4 is again due to the the front of the laser being not very well guided, thus increasing the front-view spot size of the laser. Moreover, a bunch of electrons is also injected due to the evolution of the driver in this case, with the charge \SI{15}{pC} being similar to that in the former case.

The similarities in the results of the two cases clearly highlight the usefulness of the plasma lens. For the \SI{1}{PW} laser simulated, reaching a spot size of $w_0=\SI{21.6}{\micro\metre}$ requires a focal length of about \SI{10}{m}. Employing the plasma lens, a \SI{0.86}{\milli\metre} pre-focusing distance and a \SI{4.3}{mm} long plasma lens result in similar bubble sizes as a different focusing optic with $\sim\SI{20}{m}$ focal length would.

A set of plasma lens parameters for LWFAs driven by \SI{800}{nm} wavelength lasers with powers of \SI{10}{PW} and \SI{100}{PW}, resulting from our empirical expressions and satisfying the matching condition $k_p w_2 = 2\sqrt{a_2}$ are shown in Tab.~\ref{tab:100PW}. Simulations to verify these parameter sets are, however, extremely consuming in computational resources, thus requiring fast algorithms such as the quasi-3D algorithm~\cite{ADavidsonJCP2015}.

\begin{table}
\begin{center}
  \begin{tabular}{ c | c | c | c | c }
    \hline
    $P$ [PW] & 10 & 10 & 100 & 100 \\
    $a_2$ & 4 & 4 & 4 & 4 \\
    $n_p$ [$\rm cm^{-3}$] & $2.4\times 10^{16}$ & $2.4\times 10^{16}$ & $2.4\times 10^{15}$ & $2.4\times 10^{15}$ \\
    $w_0$ [$\mu$m] & 30 & 40 & 60 & 70 \\
    $w_2$ [$\mu$m] & 136 & 136 & 431 & 431 \\
    $d$ [mm] & 35.6 & 17.7 & 694 & 563 \\
    $l$ [m] & 0.98 & 0.27 & 49 & 29 \\
    $L_d$ [m] & 6.52 & 6.52 & 206 & 206 \\
    $\tau_{\rm opt}$ [fs] & 303 & 303 & 958 & 958 \\
    $\Delta W$ [GeV] & 97.7 & 97.7 & 977 & 977 \\
    \hline
  \end{tabular}
  \caption{Plasma lens parameters for 10 PW and 100 PW laser driven LWFAs. The dephasing length $L_d$, the optimal pulse duration $\tau_{\rm opt}$ for matching the pump depletion length with the dephasing length and the energy gain $\Delta W$ for LWFAs according to Lu et al.\ are also shown~\cite{WLuPRSTAB2007}.}
  \label{tab:100PW}
\end{center}
\end{table}

\section{\label{sec:conclusion}Conclusion}

A plasma lens based on the self-focusing effect was introduced to overcome the experimental limitations for focusing high power lasers to a variety of spot sizes. With a pre-focusing optical system to focus the laser to a waist of $w_0$,  adjustable effective focal spot sizes $w_2>w_0$ can be achieved by changing the pre-plasma focusing distance $d$ and the plasma density $n_p$. Empirical formulas for $w_2$ and the plasma lens thickness $l$ were developed as Eqs.~(\ref{eq:d_w2}), (\ref{eq:zeta_empirical}) and (\ref{eq:l_d_empirical}). The focal length of the system can be reduced by the factor of $w_2/w_0$ if we compare our plasma lens with conventional focusing elements that result in the same laser spot size. Further reductions of the focal length may be achievable by cascading plasma lenses.

The upper limit of the effective focal spot size is determined by the upper limit of the pre-focusing distance $d_{\lim}$. We found the analytical correspondence $d_{\rm M}$ expressed by Eq.~(\ref{eq:d_threshould}) presents a good estimation for $d_{\lim}$ especially when $a_{0\max}$ is not large ($a_{0\max} \leq 6$). A more universal expression for $d_{\lim}$ requires further studies.

\begin{acknowledgments}
We thank the OSIRIS consortium (IST/UCLA) for access to the OSIRIS code. Furthermore, we acknowledge the grant of computing time by the J{\"u}lich Supercomputing Center on JUQUEEN under Project No. HHH23, on JUWELS under Project No. HHH45 and the use of the High-Performance Cluster (Maxwell) at DESY. This work is supported by the Helmholtz MT ARD scheme and the Helmholtz ZT-0009 project.
\end{acknowledgments}

\appendix
\section{\label{sec:cavitation}Large Cavitation Condition}
When the ponderomotive force of the laser exceeds a threshold, the plasma electrons near the laser axis are completely evacuated because the static electric field cannot balance the ponderomotive force. Naturally one may conclude that in such condition, the laser at the cavitation area behaves as if it is in vacuum. Furthermore, if the cavitation radius is larger or equal to the laser size parameter $w$, almost the whole laser beam behaves as in vacuum.

We assume the laser beam is long enough so that only the transverse ponderomotive force is taken into consideration~\cite{PMoraPOP1997}
\begin{eqnarray}
  F_p=F_{pr}=-\dfrac{1}{4\bar{\gamma}}\partial_r A^2=-\partial_r \bar{\gamma}, \label{eq:fp}
\end{eqnarray}
where the force is normalized to $m_e c \omega_p$, $A=A\left( r \right)$ is the transverse normalized vector potential profile of a linear polarized laser beam, and $\bar{\gamma}=\sqrt{1+\frac{A^2}{2}}$ is the averaged Lorentz factor for the quivering electrons. If electron cavitation does not occur, i.~e.\ electron density (normalized to plasma density $n_p$) $n_e>0$ everywhere, the transverse electrostatic force and the ponderomotive force balance each other $F_{pr}=E_r$. And from the Gauss's law $\frac{1}{r}\partial_r\left(r E_r\right)=1-n_e$, one may find the relation between the electron density and the averaged Lorentz factor~\cite{GZSunPOF1987}
\begin{eqnarray}
  1-n_e=-\dfrac{1}{r}\partial_r\left(r \partial_r \bar{\gamma}\right)=-\nabla^2_{\perp} \bar{\gamma}. \label{eq:n_e}
\end{eqnarray}
However, Eq.~(\ref{eq:n_e}) has nonphysical results when the laser ponderomotive force is stronger than the electrostatic force that an electron vacant ion column can provide. According to Gauss's law, an electron vacant ion column has the transverse electrostatic field (normalized to $m_e c \omega_p$/e)
\begin{eqnarray}
   E_r=\dfrac{1}{2}r \ {\rm for}\ r \leq r_c, \label{eq:Er}
\end{eqnarray}
where $r_c$ is the cavitation radius. If we assume the laser holds a Gaussian profile
\begin{eqnarray}
   A=a\exp\left(-\frac{r^2}{w^2}\right), \label{eq:A}
\end{eqnarray}
one may examine the cavitation occurs when
\begin{eqnarray}
   \left. \partial_r F_p \right|_{r=0} > \left. \partial_r E_r \right|_{r=0}
\end{eqnarray}
which leads to the cavitation condition
\begin{eqnarray}
   \dfrac{a^2}{w^2 \sqrt{1+\dfrac{a^2}{2}}} > \dfrac{1}{2}. \label{eq:cav_condition}
\end{eqnarray}
Particularly, under the strong relativistic condition $a^2 \gg 1$, Eq.~(\ref{eq:cav_condition}) is approximated as
\begin{eqnarray}
   \dfrac{a}{w^2} > \dfrac{1}{2\sqrt{2}}. \label{eq:cav_cond_approx}
\end{eqnarray}
Moreover, when Eq.~(\ref{eq:cav_condition}) holds, the cavitation radius $r_c$ is obtained by applying $F_p=E_r$ at $r=r_c$ to Eqs.~(\ref{eq:fp}), (\ref{eq:Er}) and (\ref{eq:A}) as
\begin{eqnarray}
   \dfrac{r_c^2}{w^2}=-\dfrac{1}{2}\ln \left[ \dfrac{w^2}{16a^2} \left( w^2+\sqrt{w^4+64} \right)\right]. \label{eq:rc}
\end{eqnarray}
We define the large cavitation condition to be $r_c \geq w$, so that almost the whole laser beam is in the electron-vacant ion column, which leads to
\begin{eqnarray}
   \dfrac{w^2}{16a^2}\left(w^2+\sqrt{w^4+64}\right)\leq \exp \left(-2\right). \label{eq:lcav}
\end{eqnarray}
This condition can be simplified if we assume $w^2 \gg 1$, thus we finally obtain the large cavitation condition
\begin{eqnarray}
   \dfrac{a}{w^2} \gtrsim 1, \label{eq:lcav_approx}
\end{eqnarray}
where $w$ is normalized to $k_p^{-1}$.

The physical idea of the large cavitation condition is similar to the upper-limit power for laser self-guiding~\cite{WMWangAPL2012}. However, in the previous paper the averaged Lorentz factor $\bar{\gamma}$ was mainly contributed by the longitudinal motion of electrons, i.~e.\ $\bar{\gamma} \propto A^2$. In our studies the transverse motion of electrons dominates, thus $\bar{\gamma} \propto A$, and the power for large cavitation to occur can be obtained by using Eq.~(\ref{eq:lcav_approx}) and $P\ \left[{\rm TW}\right] =0.0215 \times \left( a w / \lambda \right)^2$ as (in unnormalized form)
\begin{eqnarray}
   P_{\rm LC}\ \left[{\rm TW}\right] = 33.5\dfrac{n_p^2}{n_c^2}\dfrac{w^6}{\lambda^6} \label{eq:Pu}
\end{eqnarray}
where $\lambda$ is the laser wavelength and $n_c$ is the critical density for the laser.

One may notice that we use $a$ and $w$ without the subscription 0, because they can change during the propagation, and the cavitation and non-cavitation states can switch.

\begin{figure}
   \includegraphics[width=0.48\textwidth]{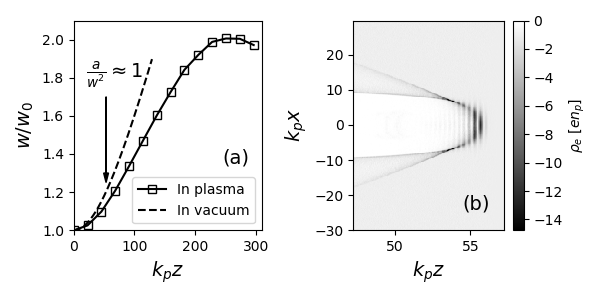}
   \caption{An example simulation showing large cavitation with $k=10$, $w_0=4$, $a_{0\max}=50$, $d=0$ and the plasma region is at $z>0$ (thus the laser is focused at the vacuum-plasma interface). (a) $w$ vs.\ z, for two cases: with plasma (square and solid line) and without plasma (dashed line). The arrow points out the approximate location where $a/w^2=1$. (b) The side slice view of the plasma electron density at where $a/w^2 \approx 1$.}
   \label{fig:l_cav}
\end{figure}

We show one example PIC simulation in Fig.~\ref{fig:l_cav}. The initial parameters are $d=0$, $a_{0\max}=50$ and $w_0=4$, with $a_{0\max}/2$ being the averaged $a_0$ (i.~e.\ $a_0=25$ in average), thus Eq.~(\ref{eq:lcav_approx}) is satisfied and large cavitation occurs at the vacuum-plasma interface. That is why for a short distance the laser size behaves similar to that in vacuum. One can observe this similarity by comparing the solid line with the dashed line in Fig.~\ref{fig:l_cav} (a). Then $a$ decreases and $w$ increases, at some distance $a/w^2=1$ appears. Although before $a/w^2=1$ occurs, the solid line does not exactly overlap with the dashed line, the location where $a/w^2=1$ is approximately the transition from ``vacuum like'' to ``plasma like'': before this point the $w$ vs.\ $z$ curve behaves similar to that in vacuum, and after this point the curve presents self-refocusing. Figure~\ref{fig:l_cav} (b) shows the electron density slice at this transition point, in which a clear large electron-vacant column can be seen.

Furthermore, if we assume that the laser pulse evolves exactly the same as in vacuum for $a/w^2>1$, a further distance should be added to $d$ in the cases of $a_1/w_1^2>1$:
\begin{eqnarray}
   d_{\rm eff}=\max \left[d,\ z_R \sqrt{\left(\dfrac{a_0}{w_0^2}\right)^{\frac{2}{3}}-1}\right], \label{eq:d_eff}
\end{eqnarray}
where $d$ is still $z_1-z_0$, and $d_{\rm eff}$ is the effective value replacing $d$ when using Eqs.~(\ref{eq:d_w2}) and (\ref{eq:l_d_empirical}). Correspondingly, $l$ also increases to its effective value
\begin{eqnarray}
   l_{\rm eff}=l+d_{\rm eff}-d, \label{eq:l_eff}
\end{eqnarray}
where $l$ is obtained by Eq.~(\ref{eq:l_d_empirical}) and $l_{\rm eff}$ is what it appears to be. These explain why in Fig.~\ref{fig:d_w2} (c) and Fig.~\ref{fig:d_vs_l} (c) the $w_2/w_0$ and $l$ are slightly larger than the fit lines for some relatively small $d$.

\section{\label{sec:prevent_self_injection}Preventing Self-Injection}
In the LWFAs, self-injection is the injection of background plasma electrons into the acceleration phase of the wake due to wave breaking or the evolution of the driver~\cite{APukhovAPB2002, CBSchroederPRE2005, SYKalmykovPOP2011, SCordeNC2013}. Self-injection is a natural injection scheme which can occur once the driver amplitude exceeds a threshold. Meanwhile, other specific injection schemes have been introduced to control the electron beam qualities, such as ionization injection~\cite{MChenJAP2006, EOzPRL2007, APakPRL2010, CMcGuffeyPRL2010, JSLiuPRL2011, BBPollockPRL2011, MChenPOP2012, MZengPOP2014}, density transition injection~\cite{HSukPRL2001, CGRGeddesPRL2008, OMartinezPRAB2017}, and optical injections~\cite{DUmstadterPRL1996, CRechatinPRL2009, RLehePRL2013}. To run LWFAs in these specific injection schemes one should avoid self-injections. In this section we discuss the lower limit of $d$ to prevent self-injection.

Under non-evolving driver and cold plasma assumptions, C.~Benedetti et al.\ have found an empirical self-injection threshold of $a_0$ using PIC simulations~\cite{CBenedettiPOP2013}
\begin{eqnarray}
   a_0^* \approx 2.75 \sqrt{1+\left(\dfrac{\gamma_0}{22}\right)^2}, \label{eq:a_th}
\end{eqnarray}
where $\gamma_0$ is the Lorentz factor for the wake phase velocity, and self-injection can occur if $a_0 > a_0^*$. We assume the wake velocity equals the group velocity of the driver, i.~e.\ $\gamma_0 = \left[1-v_g^2\right]^{-1/2} = \omega \approx k$, where $v_g$ is the laser group velocity normalized to speed of light in vacuum, $\omega$ is the laser frequency normalized to $\omega_p$ and $k$ is the laser wavenumber normalized to $k_p$. In our case, the normalized laser vector potential is varying with $z$, but its maximum value in plasma is $a_1 = a_0 / \sqrt{1+\left(d/z_R\right)^2}$. Thus the condition for preventing self-injection is $a_1 < a_0^*$ which leads to
\begin{eqnarray}
   \dfrac{d^2}{z_R^2} > \dfrac{a_0^2}{7.56\left(1+\dfrac{k^2}{484}\right)} -1. \label{eq:d_sf}
\end{eqnarray}
One may notice it is possible that the RHS of Eq.~(\ref{eq:d_sf}) is negative. In this case there is no lower limit for $d$ to prevent self-injection.

\section{\label{sec:eq_spot_size}A Derivation to the Equation of Laser Spot Size}
In this section, we give a derivation to Eqs.~(\ref{eq:energy_conserve}) and (\ref{eq:w}) using calculus of variations similar to Anderson et al.~\cite{DAndersonPOF1979}, but with a more explicit form of plasma response, i.~e.\ the RHS of Eq.~(\ref{eq:tilde_a}).

If we assume the relativistic factor of an electron is mainly contributed by the quiver motion driven by the laser, i.~e.\ $\gamma = \sqrt{1+\left|\tilde{a}\right|^2/2}$, and write down the Lagrangian
\begin{eqnarray}
   L &=& \frac{i}{2}\left(\tilde{a}\partial_z \tilde{a}^* - \tilde{a}^*\partial_z \tilde{a}\right) - \frac{1}{2k}\left| \nabla_\perp \tilde{a} \right|^2 \nonumber \\
   &&- \frac{2}{k}\left[ \left( 1+\frac{\left|\tilde{a}\right|^2}{2} \right)^{\frac{1}{2}} - \frac{\left|\tilde{a}\right|^2}{4} -1 \right], \label{eq:Lagrangian}
\end{eqnarray}
where $\tilde{a}$ is complex and $\tilde{a}^*$ is its conjugate. Thus Eq.~(\ref{eq:tilde_a}), in the case of unperturbed plasma density $n=1$, is equivalent to the extremum problem
\begin{eqnarray}
   0 &=& \delta \iint 2\pi rdr \times dz \nonumber \\
     && \times L\left( z, r; \tilde{a}, \tilde{a}^*, \partial_z \tilde{a}, \partial_z \tilde{a}^*, \nabla_\perp \tilde{a}, \nabla_\perp \tilde{a}^* \right), \label{eq:extremum}
\end{eqnarray}
where we have assumed a cylindrical symmetry, $r$ is the radial axis, and the integral is taken in the whole space.

Next we assume the laser holds the form Eq.~(\ref{eq:tilde_a_form}), where $a=a\left(z\right)$, $u=u\left(z\right)$ and $w=w\left(z\right)$. One should notice that in this section $a = \left|a\right|\exp\left(-i\varphi\right)$ is complex because it also contains the Gouy phase $\varphi\left(z\right)$. And for convenience we define
\begin{eqnarray}
   F\left(\varrho\right) = \exp\left(-\varrho^2\right), \label{eq:F_define}
\end{eqnarray}
where $\varrho \equiv r/w$,thus the partial derivatives of $\tilde{a}$ are
\begin{eqnarray}
   \partial_z \tilde{a} &=& \left(\frac{a'}{a}+iu'r^2 - \frac{r}{w}\frac{w'}{w}\frac{F'}{F}\right)\tilde{a}, \label{eq:partial_z_tilde_a} \\
   \partial_r \tilde{a} &=& \left(i2ur + \frac{1}{w}\frac{F'}{F}\right)\tilde{a}. \label{eq:partial_r_tilde_a}
\end{eqnarray}
With Eqs.~(\ref{eq:partial_z_tilde_a}) and (\ref{eq:partial_r_tilde_a}) we rewrite Eq.~(\ref{eq:Lagrangian}) as
\begin{eqnarray}
   L &=& \left[\frac{i}{2}\left(a {a^*}'-a^* a'\right) + \frac{\left|a\right|^2}{2k}\right]F^2 + \left[u' - \frac{2u^2}{k}\right]\left|a\right|^2 r^2F^2 \nonumber\\
   &&- \frac{\left|a\right|^2 F'^2}{2kw^2} - \frac{2}{k} \left(\sqrt{1+\frac{\left|a\right|^2 F^2}{2}} -1 \right). \label{eq:rewrite_Lagrangian}
\end{eqnarray}
Define a reduced Lagrangian by taking transverse integration
\begin{eqnarray}
   \mathscr{L} &\equiv& \int 2\pi rdr \times L \nonumber\\
   &=& \beta_1 \left[ \frac{i}{2}\left(a {a^*}'-a^* a'\right)w^2 + \frac{\left|a\right|^2 w^2}{2k} \right] \nonumber\\
   && + \beta_2 \left[u' - \frac{2u^2}{k}\right]\left|a\right|^2 w^4 \nonumber\\
   && - \beta_3 \frac{\left|a\right|^2}{2k} - Q \frac{2w^2}{k}, \label{eq:reduced_Lagrangian}
\end{eqnarray}
where
\begin{eqnarray}
   \beta_1 &=& \int 2\pi \varrho d\varrho \times F^2,  \label{eq:beta1}\\
   \beta_2 &=& \int 2\pi \varrho d\varrho \times \varrho^2 F^2,  \label{eq:beta2}\\
   \beta_3 &=& \int 2\pi \varrho d\varrho \times F'^2,  \label{eq:beta3}\\
   Q &=& \int 2\pi \varrho d\varrho \times \left(\sqrt{1+\frac{\left|a\right|^2 F^2}{2}} -1\right), \label{eq:Q} \label{eq:beta123Q}
\end{eqnarray}
thus Eq.~(\ref{eq:extremum}) is reduced to
\begin{eqnarray}
   0 &=& \delta \int dz \times \mathscr{L}\left( z; a, a^*, u, w, a', {a^*}', u', w'\right). \label{eq:reduced_extremum}
\end{eqnarray}
With the Euler-Lagrange equation we obtain
\begin{eqnarray}
   \frac{\partial\mathscr{L}}{\partial a} &-& \frac{d}{dz}\frac{\partial\mathscr{L}}{\partial a'} = 0: \nonumber\\
   0 &=& - \frac{2w^2}{k}\partial_a Q + \beta_1 \left( i{a^*}'w^2 + i a^* w w' + \frac{a^* w^2}{2k}\right) \nonumber\\
   && + \beta_2 \left(u'-\frac{2u^2}{k}\right)a^*w^4 - \beta_3 \frac{a^*}{2k} , \label{eq:Euler-Lagrange_a} \\
   \frac{\partial\mathscr{L}}{\partial u} &-& \frac{d}{dz}\frac{\partial\mathscr{L}}{\partial u'} = 0: \nonumber\\
   0 &=& - \frac{4u}{k}\left|a\right|^2 w^4 - \left(\left|a\right|^2 w^4\right)', \label{eq:Euler-Lagrange_u} \\
   \frac{\partial\mathscr{L}}{\partial w} &-& \frac{d}{dz}\frac{\partial\mathscr{L}}{\partial w'} = 0: \nonumber\\
   0 &=& \beta_1 \left[ i\left( a {a^*}'-a^* a' \right) + \frac{\left|a\right|^2}{k} \right]w \nonumber\\
   && + 4\beta_2 \left(u' - \frac{2u^2}{k}\right)\left|a\right|^2 w^3 - 4Q\frac{w}{k}. \label{eq:Euler-Lagrange_w}
\end{eqnarray}
By multiplying $a$ on both sides of Eq.~(\ref{eq:Euler-Lagrange_a}) and taking its imaginary part we recover Eq.~(\ref{eq:energy_conserve}) (notice $a\partial_a Q$ is real); by taking its real part and comparing with Eq.~(\ref{eq:Euler-Lagrange_w}) we have
\begin{eqnarray}
   \beta_2 \left(\frac{u'}{2} - \frac{u^2}{k}\right) +  \frac{\beta_3}{4kw^4} = \frac{Q - a \partial_a Q}{k\left|a\right|^2 w^2}. \label{eq:temp_eq}
\end{eqnarray}
From Eqs.~(\ref{eq:energy_conserve}) and (\ref{eq:Euler-Lagrange_u}) we obtain
\begin{eqnarray}
   -\frac{2u}{k} = \frac{w'}{w} \label{eq:u_w_prime}
\end{eqnarray}
which also suggests $k/2u$ is the radius of the wavefront curvature. Thus
\begin{eqnarray}
   w'' = \frac{\beta_3}{\beta_2 k^2 w^3}\left[1 - \frac{4w^2 \left(Q - a \partial_a Q\right)}{\beta_3\left|a\right|^2}\right]. \label{eq:w_prime_prime}
\end{eqnarray}
By inserting Eq.~(\ref{eq:F_define}) in Eq.~(\ref{eq:beta1}) - (\ref{eq:Q}) one can obtain $\beta_2 = \pi/4$, $\beta_3 = \pi$ and $\partial_a Q - Q = \pi \left|a\right|^4 /128 + \mathcal{O}\left(\left|a\right|^6\right)$. Finally Eq.~(\ref{eq:w}) is recovered.

One may notice the assumption that the laser holds the form Eq.~(\ref{eq:tilde_a_form}) may not be correct, i.~e.\ the laser may not perfectly maintain a Gaussian profile after propagating a while in plasma. In this case to define a laser radius $w$ is arbitrary. Thus Eq.~(\ref{eq:w}) is just an approximate function for the laser spot size evolution in plasma.

\bibliography{self_refocusing.bib}{}

\end{document}